\documentclass[AMA,STIX1COL]{WileyNJD-v2}
\usepackage{moreverb}
\usepackage{amsthm,url,amsmath}
\usepackage{float}
\usepackage{tikz}
\usepackage{tikz}
\usepackage{wasysym}
\usetikzlibrary{calc,arrows,positioning,shapes,shapes.gates.logic.US,trees}
\usepackage{threeparttable}
\usepackage{siunitx} % Required for alignment
\usepackage{multirow} % Required for multirows
\usepackage{booktabs} % For prettier tables
\usepackage{longtable} % To display tables on several pages
\usepackage{rotating} % To display tables in landscape

\newcommand{\transpose}[1]{\ensuremath{#1^{\scriptscriptstyle T}}}
\newcommand\BibTeX{{\rmfamily B\kern-.05em \textsc{i\kern-.025em b}\kern-.08em
T\kern-.1667em\lower.7ex\hbox{E}\kern-.125emX}}

\articletype{Research Article}%

\begin{document}
\title{Application of Cox model to predict the survival of patients with Chronic Heart Failure (CHF):  \itshape{A latent Class regression approach}}

\author[1,2,3]{John L. Mbotwa*}

\author[2]{Marc de Kamps}

\author[1]{Paul D. Baxter}

\author[1,4]{Mark S. Gilthorpe}

\authormark{MBOTWA \textsc{et al}}

\address[1]{\orgdiv{Leeds Institute for Data Analytics}, \orgname{University of Leeds}, \orgaddress{\state{LS$2$ $9$NL}, \country{UK}}}

\address[2]{\orgdiv{Institute of Artificial and Biological Intelligence}, \orgname{University of Leeds}, \orgaddress{\state{LS$2$ $9$JT}, \country{UK}}}

\address[3]{\orgdiv{Department of Applied Studies}, \orgname{Malawi University of Science and Technology}, \orgaddress{\state{P.O. Box $5196$}, \country{Malawi}}}

\address[4]{\orgdiv{The Alan Turing Institute}, \orgname{British Library}, \orgaddress{\state{ 96 Euston Road, London}, \country{UK}}}

%\address[3]{\orgdiv{Org Division}, \orgname{Org name}, \orgaddress{\state{State name}, \country{Country name}}}

\corres{*John Lenard Mbotwa, Leeds Institute for Data Analytics, Worsley Building, Clarendon Way, Leeds, LS$2$ $9$NL, UK. \email{J.L.Mbotwa@leeds.ac.uk}}

%\presentaddress{Present address}

\abstract[Abstract]{Most prediction models that are used in medical research fail to accurately predict health outcomes due to methodological limitations. Using  routinely collected patient data, we explore the use of a Cox proportional hazard (PH) model within a latent class framework to model survival of patients with chronic heart failure (CHF). We identify subgroups of patients based on their risk with the aid of available covariates. We allow each subgroup to have its own risk model. We choose an optimum number of classes based on the reported Bayesian information criteria (BIC). We assess the discriminative ability of the chosen model using an area under the receiver operating characteristic curve (AUC) for all the cross-validated and bootstrapped samples. We conduct a simulation study to compare the predictive performance of  our models. Our proposed latent class model outperforms the standard one class Cox PH model.}

\keywords{Latent class model, Cox proportional hazards model, Cardiovascular, survival analysis, prediction, heterogeneity}

\jnlcitation{\cname{%
\author{Mbotwa J.}, 
\author{de Kamps M.}, 
\author{Baxter PD},  and 
\author{Gilthorpe MS}} (\cyear{2019}), 
\ctitle{Application of Cox model to predict the survival of patients with Chronic Heart Failure (CHF):  \itshape{A latent Class regression approach}}, \cjournal{Statistics in Medicine}, \cvol{2019;00:0--0}.}

\maketitle

%\footnotetext{\textbf{Abbreviations:} ANA, anti-nuclear antibodies; APC, antigen-presenting cells; IRF, interferon regulatory factor}

\section{Introduction}\label{sec1}
Chronic Heart Failure (CHF) is one of the leading causes of hospitalizations and deaths more especially in older people, and this leads to a substantial clinical and economic burden to the government{\cite{R1}}. Prediction models are indisputably useful as they complement clinical decisions in modern medicine \cite{R2}. For instance, using risk prediction models to accurately understand the dynamics of survival patterns amongst patients with CHF conditions would provide guidance to health care professionals in decision making on how to improve the delivery of care.  These models also help in reducing unnecessary costs because resources are distributed to targeted subgroups of patients{\cite{R3}}. It is therefore imperative that these models should provide accurate predictions to fully guide health care professionals in decision making on how to improve prediction of patient outcomes and reduce unnecessary costs of care \cite{R3}.  \\
Most of the risk prediction models used in cardiovascular studies fail to make accurate predictions due to both constraints in data quality and methodological limitations \cite{R4}. Data quality is mostly compromised due to missing data. Missing data also affects the estimation of model parameters and increases the risk to bias \cite{R25}. As a result, the problem of missing data has a significant effect on the conclusions from prediction models. These models also perform poorly when predicting narrowly targeted subgroups of patients because they are primarily developed to make population-level predictions. Suppose a group of patients with CHF are being followed-up for a fixed period, it is very likely that some patients will have a higher likelihood of dying during the follow-up period while others are likely not to die up to the end of the study. In this situation, it is important to identify meaningful subgroups (i.e.\ latent classes) of patients with similar survival patterns based on the available covariates. 
To understand this, we require adequate data on several putatively causal factors and patient characteristics that help reflect the underlying true description of the study population. The effect of these factors and patient characteristics on survival may vary across the latent subgroups during the follow-up period. \\ 
In this analysis we propose to use latent class regression analysis (LCRA) in which covariates are allowed not only to predict the outcome (e.g.\ survival patterns) in each class but also to predict class membership. Patients are probabilistically assigned into subgroups known as classes. These classes are designed in such a way that within each class patients have similar characteristics while between classes patients are different from each other. Latent classes contain different proportions of the study population and the relationship between the outcome of interest and the covariates may differ across classes\cite{R5}.The practical utility of using LCRA is that these latent classes may correspond to a specific subgroup of patients with specific characteristics that predict their survival. The rationale is that one class will be more susceptible to deaths compared to another class hence each class is assigned its own risk model.
\section{Methods}
\subsection{Study design, data collection and ethics statement}
In this analysis we used data from the United Kingdom Heart Failure Evaluation and Assessment of Risk Trial 2 (UK-HEART 2) which is a prospective cohort of ambulant patients with signs and symptoms for CHF\cite{R11}. The UK-HEART 2 cohort recruited 1802 patients with CHF who attended specialist cardiology clinics in four UK hospitals between July 2006 and December 2014\cite{R9}. Patients were recruited in the study if they had clinical signs and symptoms of CHF conditions present for at least 3 months, were at least 18 years old  and had left ventricular ejection fraction less than or equal to $45\%$ \cite{R9,R11}. The study protocol was approved by the ethics committees at each of the participating institutions\cite{R10}, and all patients gave an informed consent to take part in the study\cite{R11}.  A more detailed description of the cohort and the inclusion exclusion criteria has been reported elsewhere\cite{R9,R10,R11}.
\subsection{Statistical methods}
Latent class regression analysis is a single-level regression analysis procedure in which data is split into subgroups or latent classes. Each subgroup contains its own risk model so that the estimated model parameters are specific to that class\cite{R8}. The optimum number of latent classes to be formed is typically determined by examining the Bayesian information criteria (BIC) as this has been shown to outperform other model fit statistics under simulations \cite{R18}. The ideal strategy for determining the number of classes may also be driven by interpretability decisions such as having clinical utility \cite{R6,R7,R8} and therefore needs to be considered and developed depending on context and application. The probability of any patient belonging to a particular class is based on the similarities in characteristics of patients attributed to each class. Individuals may be probabilistically assigned to more than one class, with everyone's total assignment over all classes summing to one.   \\ 
One challenge with the latent class modelling approach is the sensitivity to starting values. The starting values are used to maximize the likelihood function when estimating model parameters. If the starting values are far from the optimum solution, the likelihood function fails to converge or takes longer to do so. For example, if $30$ random starts are used, sometimes only $15$ of them may give meaningful solutions when the likelihood function is maximized. For a solution to be meaningful, we expect the highest likelihood value to be replicated many times. When this fails, it simply means that the solution has not been achieved and one needs to increase the random starts to converge on a global optimum solution. The values that gave an optimum likelihood can be used as initial values for the final model in-order to reduce the search process \cite{R17}. 
\subsubsection{The General Latent Class model :}
The Latent Class Model (LCM) comes from a family of finite mixture models which classifies observations into classes to model unobserved heterogeneity within a population.  Suppose that a population $P$ is naturally partitioned into $g$ classes $p_1, p_2, \ldots, p_g$. Let $\boldsymbol y$ be an outcome variable from $g$ distinct classes. Let the probability density functions for each of the $g$ classes be $f_1, f_2,  \ldots, f_g$ with corresponding proportions $\pi_1, \pi_2, \ldots, \pi_g$   for belonging to any of the respective classes. Thus the mixture density function of $\mathbf{y}$ is defined as 
\begin{equation}\label{eq:1}
\displaystyle{f(\boldsymbol y|\boldsymbol {z, x}, \lambda)=\sum_{i=1}^{g} \pi_i( \boldsymbol{z}| \gamma_i,\boldsymbol \delta_i) f_i(\boldsymbol{y}|\boldsymbol{x},\boldsymbol\beta_i)}
\end{equation}
where $\lambda =(\gamma_i,\boldsymbol{\delta_i},\boldsymbol{\beta_i})$ represents a collection of model parameters and  $\pi_i(\boldsymbol{z}| \gamma_i,\boldsymbol {\delta_i})$'s are class-membership probabilities that are estimated for each class and are dependent on a vector of covariates $\boldsymbol{z}$ such that 

\begin{equation}
\sum_{i=1}^g \pi_i(\boldsymbol{z}|\boldsymbol \gamma_i,\boldsymbol\delta_i)=1
\end{equation}
with  $0\leq\pi_i(\boldsymbol{z}| \gamma_i,\boldsymbol\delta_i)\leq 1$ and $f_i(\boldsymbol{y}|\boldsymbol{x},\boldsymbol \beta_i)$ is the conditional probability density function for the observed response in the $ith$ class model and $\boldsymbol{x}$  is the covariate vector.\\
For a class membership model, the structural part of the model is given by 
\begin{equation}
\text{logit}(\pi_i(\boldsymbol{z}| \gamma_i,\boldsymbol{\delta_i}))= \gamma_i+ \boldsymbol{z}\transpose{\boldsymbol\delta_i} 
\end{equation}
hence 
\begin{equation}
\pi_i(\boldsymbol{z}| \gamma_i,\boldsymbol{\delta_i})=\frac{\exp(\gamma_i+ \boldsymbol{z}\transpose{\boldsymbol{\delta_i}})}{\sum_{j=1}^g{\exp(\gamma_j+ \boldsymbol{z}\transpose{\boldsymbol\delta_j})}}
\end{equation}
where  $\boldsymbol{z}$ is a $(p \times 1)$ covariate vector for the class-membership model and $\transpose{\boldsymbol {\delta_i}}$ is the transpose of the vector 
$\boldsymbol {\delta_i}$ for the multinomial logistic class-membership model. Suffice to say, covariate vectors $\boldsymbol{x}$ and $\boldsymbol{z}$ do not necessarily have to be the same. \\
In this paper, we apply survival analysis within a latent class framework described in equation \ref{eq:1} to predict subgroups of patients with different prognosis based on the available covariates. This is in conjunction with the prediction of survival distributions for different subgroups of patients using patient covariates. The distribution of the survival time variable for each component in equation \ref{eq:1} can be parametric (i.e.\ a scenario with distributional assumptions about the survival times), semi-parametric (i.e.\ a scenario with relaxed distributional assumptions) or non-parametric (i.e \ a scenario without distribution assumptions about the survival times). If we assume a parametric model for the response variable, the component's densities are assumed to be from the same family. Some common distribution functions that may be considered appropriate for survival times in a parametric case include the exponential, Gamma and Weibull \cite{R27}. In a semi-parametric case, the Cox  proportional hazard model is an example. \\ \\
If $T$ is a non-negative random variable representing time-to-death or time-to-loss-of-follow-up or simply time to the end of the study for all patients with CHF, and suppose that for each individual we have a $k \times 1$ covariate vector, denoted $\boldsymbol{x}$ that  affects the survival of patients in each class, we can define our survival model within a Latent class framework as follows:
\begin{equation}\label{eq:2}
\displaystyle{S(\boldsymbol{t|x,z, \theta})=\sum_{i=1}^{g} \pi_i(\boldsymbol z| \gamma_i,\boldsymbol\delta_i) S_i(\boldsymbol{t}|\boldsymbol{x, \beta_i)}},
\end{equation}
where $\boldsymbol \theta=(\gamma_i,\boldsymbol{\delta_i}, \boldsymbol \beta_i)$ is the collection of parameters and $\pi_i(\boldsymbol z| \gamma_i,\boldsymbol\delta_i)$ satisfies the constraints in \ref{eq:1}. The vectors $\boldsymbol{x}$ and $\boldsymbol{z}$  may include patient characteristics and medications. These covariates do not necessarily need to be the same in each class.  
If the effects of the covariates on the hazards (i.e.\ the instantaneous risk of event) in each class is constant during the entire duration of the follow-up period, then the hazard function can be specified as: 
\begin{equation}\label{eq:3}
\displaystyle{h_i(\boldsymbol {t}|\boldsymbol x,\boldsymbol \beta_i)=h_{0i}(\boldsymbol{t})\exp({\boldsymbol x\transpose{\boldsymbol \beta_i} })},
\end{equation}
where $h_{0i}(\boldsymbol{t})$ is the baseline hazard for class $i$ and $\exp({\boldsymbol x\transpose{\boldsymbol \beta_i} })$ is the relative risk associated with a vector of predictors $\boldsymbol x$. We can derive a survival function from equation \ref{eq:3} as follows:
\begin{equation}\label{eq:4}
\displaystyle{S_i(\boldsymbol{t}|\boldsymbol{x},\boldsymbol{\beta_i})= \Big[S_{0i}(\boldsymbol{t})\Big]^{\exp({\boldsymbol {x} \transpose{\boldsymbol {\beta_i}} })}}
\end{equation}
where
\begin{equation}\label{eq:5}
\displaystyle{S_{0i}(\boldsymbol{t})=\exp\Big\{-\int_0^t{h_i(\boldsymbol{u}|\boldsymbol{x,\boldsymbol\beta_i)\,\mathrm{d}u}}\Big\}}
\end{equation}
is the baseline survival for class $i$ at time $t$ given a vector of predictors $\boldsymbol{x}$ in that class. The baseline hazard $h_{0i}(\boldsymbol{t})$ in equation \ref{eq:3} is assumed to be an unknown arbitrary non-negative function of time{\cite{R15,R16}}. The only parametric part of the model in equation \ref{eq:4} is $\exp({\boldsymbol x\transpose{\boldsymbol \beta_i} })$. The maximum likelihood procedure fails to estimate parameters for the likelihood function of equation \ref{eq:3} accurately because the baseline hazard function is not assumed to take any particular form. These parameters can be estimated using the partial-likelihood approach {\cite{R16}}. The partial-likelihood function is derived by taking the product of the conditional risk at time $t_i$ given the set of individuals who have not failed or been  censored by that time.
\subsubsection{Model Selection}
Table \ref{tab:summary} presents summary statistics on the key variables used in this analysis.  
\begin{table} [th]
\caption {Summary statistics for the key variables used in the models} \label{tab:summary} 
\centering
\setlength\belowcaptionskip{-3ex}
\small\sf
\begin{tabular}{ l  l  l  l} 
\toprule
${}$ & $\underline{\textbf{1-class Model}}$  & \quad \quad \quad \quad \quad \quad \quad \quad $\underline{\textbf{2-class Model}}$ \\ \\
$\textbf{Variable}$ & $\textbf{Overall (1796)}$ & \quad \quad $\textbf{Class 1 (1566)}$ & $\textbf{Class 2 (230)}$   \\  \midrule
$\text{Survival time (yrs)}$ & $5.43$ $(3.23, 7.87)$ & \quad \quad$5.38$ $(3.23, 7.87)$ & ${}$   \\  
$\text{Haemoglobin (g/dl)}$ & $13.46$ $(1.78)$ & \quad \quad $13.80$ $(1.58)$ & $11.14$ $(1.18)$   \\  
$\text{Age (yrs)}$ & $69.65 $ $(12.52)$ & \quad \quad $69.22$ $(12.69)$ & $72.5$ $(10.94)$   \\ 
$\text{Diabetes}$ & $504$ $(28.1\%)$ & \quad \quad $368$ $(23.5 \%)$ & $136$ $(59.1 \%)$  \\ 
$\text{Sex (males)}$ & $1313$ $(73.1\%)$ & \quad \quad $1160$ $(74.1 \%)$ & $153$ $(66.5 \%)$  \\ 
$\text{dead}$ & $1061$ $(59.1\%)$ & \quad \quad $1046$ $(66.8 \%)$ & $15$ $(6.5\%)$  \\ 
\bottomrule
\end{tabular} \\ [8pt]
\begin{tablenotes}
\centering
      \scriptsize
 \item  \textbf{${}^\star$ Median (interquartile range) for survival time} \\
\item  \textbf{${}^\star$ Mean (standard deviation) for haemoglobin status and age} \\ 
\item  \textbf{${}^\star$  n(\%) for sex, diabetes and death status}
    \end{tablenotes}
\end{table} \\
%\begin{table}[htbp]
%\small \sf \caption {Summary statistics for the key variables used in the models} \label{tab:summary}
% % \centering
%  \sisetup{table-format=1.4, table-number-alignment=center}
%  \begin{tabular}{S[table-format=3.2] *{2}{S}S[table-format=3.2] }
%    \toprule & {1-class Model} &{2-class Model} \\
%    \cmidrule(lr){2-2} \cmidrule(lr){3-3}
%   {$\textbf{Variable}$} & {$\textbf{Overall (1796)}$} & {$\textbf{Class 1(1566)}$} & {$\textbf{Class 2 (230)}$}  \\  
%    \midrule
%   { \text{Survival time(yrs)}} & {3.40(2.11, 5.78)} & {3.86 (2.41,5.89)} & {1.13 (0.50,2.27)}   \\
%    {\text{Haemoglobin}}        & {13.46 (178)}      & {13.80 (1.58)} & {11.14 (1.18)}   \\
%    {\text{Age (yrs)}}          & {69.65 (12.52)}    & {69.22 (12.69)} & {72.5 (10.94)}   \\
%    {\text{Diabetes}}           & {504 (28.1\%)}     & {368(23.5 \%)} & {136(59.1\%)}   \\
%    {\text{Sex (Males)}}        & {1313 (73.1\%)}    & {1160(74.1 \%)} & {153(66.5\%)}   \\
%     {\text{Dead}}              & {1061 (59.1 \%)} & {1046(66.8 \%)} & {15(6.5\%)}   \\
%    \bottomrule
%  \end{tabular} %\\[10pt]
%  \begin{tablenotes}
%%\centering
%      \scriptsize
%     \item  \textbf{${}^\star$ Median (interquartile range) for survival time, mean (standard deviation) for haemoglobin status and age. \\
%     \item  \textbf{${}^\star$  n(\%) for sex, diabetes and death status.
%\end{table} 
\noindent 
In a two-class model, $1566$ $(87.2\%)$ were allocated to Class $1$ while $230$ $(12.8\%)$ were allocated to Class $2$. As Table \ref{tab:summary} suggests, a large proportion of patients in Class $2$ had diabetes $136$ $(59.1 \%)$ and lower levels of haemoglobin (mean=$11.1$g/dl) compared to Class $1$ which had only $23.5\%$ patients with diabetes and an average haemoglobin of $13.8$g/dl. Class $1$ was also dominated by males $(74.1 \%)$ and more deaths were recorded in Class $1$  $(66.8 \%)$. In terms of age, Class $1$ had a slightly lower average age of $69.2$yrs compared to Class $2$ which had an average age of $72.5$yrs. \\ \\
We examined the one-, two- , three-, four- and five-class models to determine an optimum number of classes. Due to a limited dataset, selection of model covariates was based on an all subset regression procedure. All candidate explanatory variables were included in the model if they were significant predictors of either class or survival. Ideally, we propose a lifecourse setting in which early life exposures are significantly associated with subgroups of patients. Similarly, midway exposures are either predictors of subgroups or the final outcome (e.g.\ death or survival). Later life exposures only affect the final outcome since chrystalization happens before the final outcome.  The first model we examined was the global (one class) model in which a single model was used to predict survival for the whole population. In our second model, we assumed that the underlying population was heterogeneous hence we split our data into two subgroups based on patient covariates. We fitted a separate Cox proportional hazard model for each subgroup. The same approach was repeated for three, four and five class models. We noted that the higher the number of latent subgroups, the more convergence problems we encountered. One possible reason is that, any increase in the number of latent classes results in a corresponding increase in the number of model parameters. \\ \\
In this paper, we focus our discussion on the first three models. In the one-class model, a single model was used to predict survival for the whole population. In each component (i.e.\ subgroup) identified using a two- or three-class model, survival rates were predicted by a separate Cox proportional hazard model.  We extracted some model fit statistics as summarised in Table \ref{tab:fitstats}. From the three models we inspected, different model fit indexes favoured different models. The Alkaike Information Criteria (AIC) and $-2$Log-Likelihood statistics improved with an increase in the number of classes. The three-class model was therefore favoured in terms of AIC and $-2$Log-Likelihood, whereas the Bayesian Information Criteria (BIC) improved from a one-class model to a two-class model and then deteriorated for a three-class model. The result of sample-size adjusted BIC is consistent with BIC, indicating the model with two-classes is better. For the sake of interpretability and clinical utility, we also favour fewer classes and select the model with the smallest BIC and aBIC, as recommended in most simulation studies (see for example Nylund et al\cite{R18}). 
%\begin{table}[H]
%\small \sf  \caption {Model fit statistics for the three models} \label{tab:fitstats} 
%\vspace*{0cm}
%\begin{tabular}{ l  l  l l}
%\toprule
%%\hline
%$\textbf{Fit Index}$ & $\textbf{One-class}$ & $\textbf{Two-class}$ & $\textbf{Three-class}$  \\  
%\midrule
%$-2\text{Log-Likelihood}$ & $12383.81$ & $12258.39$ & $12235.15$  \\  
%$\text{Number of Parameters}$ & $3$ & $10$  & $17$   \\  
%$\text{Alkaike Information Criteria}$ & $12389.81$ & $12278.39$ & $12269.15$   \\ 
%$\text{Bayesian Information Criteria}$ & $12406.30$ & $12333.32$ & $12362.54$  \\ 
%\bottomrule
%\end{tabular} \\[10pt]
%\end{table} 
\begin{table}[th]%
\small \sf  \caption {Model Fit statistics for the three models} \label{tab:fitstats}
  \centering
  \sisetup{table-format=1.4, table-number-alignment=center}
  \begin{tabular}{S[table-format=1.3] *{2}{S}S[table-format=1.2] }
    %\toprule & {sub-problem 1} &{sub-problem 2}\\
    \toprule
    %\cmidrule(lr){2-2} \cmidrule(lr){3-3}
   {$\textbf{Fit Index}$} & {$\textbf{One-class}$} & {$\textbf{Two-class}$} & {$\textbf{Three-class}$}  \\  
    \midrule
   { -2\text{LL}} & 12383.81 & 12258.39 & 12235.15  \\
    {\text{N}} &  3 & 10  & 17   \\
    {\text{AIC}} & 12389.81 & 12278.39 & 12269.15   \\
    {\text{BIC}} & 12406.30 & 12333.32 & 12362.54  \\ 
 {\text{aBIC}} & 12396.77 & 12301.55 & 12308.53  \\ 
    \bottomrule
  \end{tabular} %\\[12pt]
  \begin{tablenotes}%%[341pt]
\centering
      \scriptsize
     \item  \textbf{${}^\star$ -2LL= -2Log-Likelihood; ${}^\star$ N= Number of parameters} \\
     \item  \textbf{${}^\star$ AIC =Alkaike Information Criteria; ${}^\star$ BIC = Bayesian Information Criteria \\
     ${}^\star$ aBIC =adjusted Bayesian Information Criteria}
    \end{tablenotes}
\end{table}
\subsubsection{Model performance evaluation and Validation}
\begin{figure}[th]
  \centering
    \includegraphics[scale=0.5]{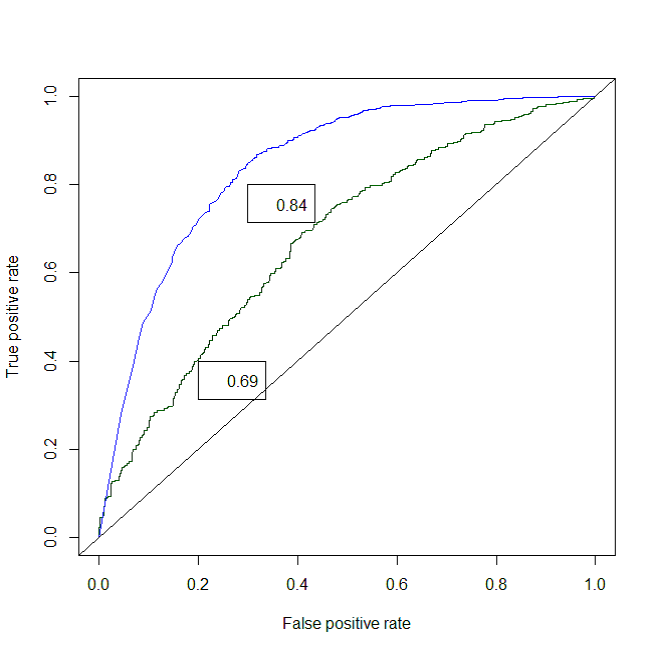}
    \caption{A ROC curve for assessing the predictive ability for the one-class and two-class Models. The curve closer to the straight line (green) is for the one-class model. The curve for the two-class model (blue) is further from the straight line.} 
    \label{fig:Areaunderthecurve1}
 \end{figure}

For a one-class model, we included age (i.e.\ age at recruitment), haemoglobin (i.e.\ the blood haemoglobin concentration in g per dl) and diabetes status (i.e.\ history of diabetes) as covariates. These covariates were identified as important joint predictors of survival using the UK-HEART study data. In a two-class model, the same set of variables were used as predictors of survival. Additionally, we identified patient's diabetes status and haemoglobin levels as important class predictors. Age was not found to be an important class predictor in this study. We used the area under the receiver operating characteristic curve (AUC) to assess the predictive ability of our models. The AUC has been widely used in medical research to assess the diagnostic ability of a biomarker to discriminate between the diseased and health subjects \cite{R19,R20,R24}. In our study we used AUC to quantify how well our model is able to discriminate between the two subgroups, e.g.\ high risk and low risk latent classes. An area under the ROC curve can yield values ranging from $0.5$ to $1$. A minimum value of $0.5$ shows the result is equivalent to chance (i.e.\ random guess). A value above $0.8$ indicates that a model has a high predictive capacity to discriminate between the hidden groups. A value of $1$ shows that model prediction is perfect. We compared the one-class Cox proportional hazard model against the two-class Cox proportional hazard model in terms of AUC. A two-class model yielded an improved prediction of survival patterns amongst patients with CHF compared to a one-class model. An AUC of $0.84$ was achieved in a two-class model compared to $0.69$ for the one-class model as shown in Figure~\ref{fig:Areaunderthecurve1}. Our proposed modelling approach therefore outperformed the traditional one-class model in terms of predicting survival of patients with CHF. \\ 
%\newpage 
\noindent
We implemented a $k$-fold cross-validation procedure to validate our best fitting model as recommended by Grimm, Mazza and Davoudzadeh \cite{R13}.  To implement a $k$-fold cross-validation we  first divide our data into $k$ roughly equal sized partitions. The first $k-1$ partitions are used as a training set and a model is tested and validated using the $k$th partition. The procedure is repeated until a model is trained and tested  $k$ times. \\ 
We examined the validity of the standard one-class model and the two-class model using a $10$-fold cross-validation technique. We randomly divided our data into $10$ approximately equal non-overlapping partitions. We estimated our models in each case using nine of the partitions which formed a training dataset. A model obtained from the training dataset was then applied to the $10\text{th}$ partition (test dataset) while fixing the parameters at the values obtained from the training data. The AUC was calculated for each of the $10$ test samples. The mean AUC for the one-class  model using the $10$ test samples was $0.69$, which is equivalent to the initial $0.69$. The mean AUC for the two-class model using the $10$ test samples was $0.79$, which is clearly lower than the initial  $0.84$; we expect a smaller sample to yield a lower AUC due to larger standard errors and thus poorer precision. However, some AUC values from the $10$ validation samples gave larger values, which could just be due to chance. To confirm the results, we also examined the predictive capacity of our models by using a bootstrap re-sampling procedure. The bootstrap re-sampling method is the process that involves creating datasets from the original dataset without making any further assumptions\cite{R26}. Estimates of interest, such as the AUC can be calculated for each bootstrap sample and the distributions inspected. We drew $30$ random bootstrap samples and fitted both one-class and two-class models in each case.  We calculated AUC estimates for each model using the bootstrap samples. The mean AUC for the one-class and two-class Cox proportional hazard models across the $30$ bootstrap samples was $0.69$ and $0.75$ respectively. The results from the two cross-validation procedures were consistent as summarised in Table {\ref{tab:validation}}. \\ \\
\begin{table} [th]
\small\sf
\caption {Summary statistics for a Model validation} \label{tab:validation} 
\centering
\setlength\belowcaptionskip{-3ex}
\small\sf
\begin{tabular}{ l  l  l  l}
\toprule
${}$ &  \quad \quad \quad \quad  & $\underline{\textbf{1-class Model}}$  & \quad $\underline{\textbf{2-class Model}}$ \\ \\
$\textbf{Method}$ & $\textbf{N}$ & $\textbf{Mean AUC (95\% CI)}$   \quad \quad \quad &
 $\textbf{Mean    AUC (95\% CI)}$   \\  \midrule
$\text{k-fold cross-validation}$ & $10$ & $0.69$ $(0.66-0.72)$  & $0.79$ $(0.70-0.87)$  \\ 
$\text{Bootstrap}$ & $30$  & $0.69$ $(0.69-0.70)$ & $0.75$ $(0.71-0.79)$  \\  
\bottomrule
\end{tabular} %\\[12pt]
\begin{tablenotes}
\centering
      \scriptsize
 \item \textbf{${}^\star$  AUC= Area under the Curve} \\
\item \textbf{${}^\star$ IQR= Inter Quartile Range} \\
\item \textbf{${}^\star$ SD =Standard deviation}
    \end{tablenotes}
\end{table} 

%\newpage
\noindent
The cross-validation and bootstrap procedure was done in R, but Latent Class Models were done in M\textit{plus}. We used the M\textit{plus} automation package developed by Hallquist and Willey \cite{R14} to run models in M\textit{plus} from R. An Mplus script for the two-class model was created and applied to the training dataset. Using an M\textit{plus} automation package another M\textit{plus} script was developed with parameters fixed at those obtained from the training dataset \cite{R13}. 
\section{Simulation design}
To evaluate the performance of the prediction accuracy of the two modelling frameworks, we designed a simulation study. The simulation was designed to assess the role of predictors in the context of lifecourse exposures predicting a later-life health outcome. 
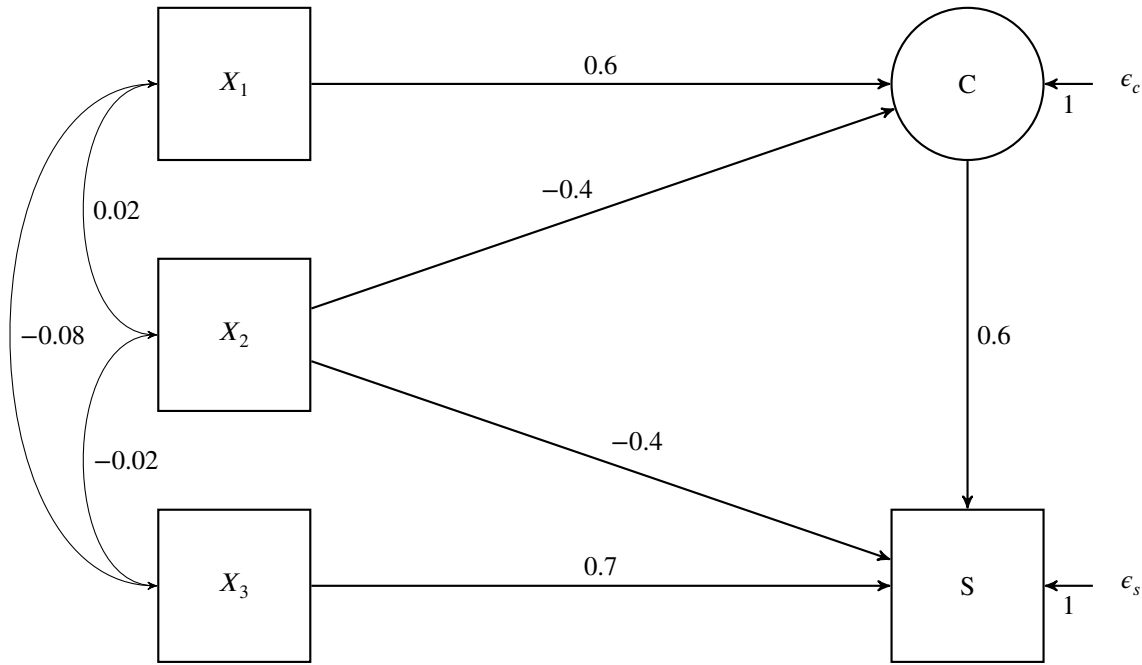
\begin{figure}[th]
\begin{tikzpicture}[auto,scale=3,
latent/.style={circle,draw,thick,inner sep=0pt,minimum size=20mm},
error/.style={circle,inner sep=0pt,minimum size=10mm},
manifest/.style={rectangle,draw,thick,inner sep=0pt,minimum size=20mm},
corr/.style={<->,>=stealth', bend left=270},
intercept/.style={regular polygon,
regular polygon sides=3,draw,thick,inner sep=0pt,
minimum size=10mm},
mean/.style={regular polygon,regular polygon sides=3,draw,thick,inner
sep=0pt,minimum size=10mm},
paths/.style={->, thick, >=stealth'},
variance/.style={<->, thick, >=stealth', bend left=270},
varianceTop/.style={<->, thick, >=stealth', bend right=270, looseness=2},
unique/.style={<->, thick, >=stealth', loop above=270, looseness=8},
factvar/.style={<->, thick, >=stealth', loop left=270, looseness=8}
]
\tikzset{mystyle/.style={->,double=black}}
\node [manifest] (x1) {$X_1$};
\node [manifest] (x2) [below = 0.5in of x1] {$X_2$};
\node [manifest] (x3) [below = 0.5in of x2] {$X_3$};
\node [latent] (c) [right = 3in of x1] {C};
\node [manifest] (s) [right = 3in of x3] {S};
\node [error] (e1) [right = 0.25in of c] {$\epsilon_c$};
\node [error] (e2) [right = 0.25in of s] {$\epsilon_s$};
\draw [paths] (x1) to node {$0.6$} (c);
\draw [paths] (x2) to node {$-0.4$} (c);
\draw [paths] (x2) to node {$-0.4$} (s);
\draw [paths] (x3) to node {$0.7$} (s);
\draw [paths] (c) to node {$0.6$} (s);
\draw [corr] (x1) to node {$0.02$} (x2);
\draw [corr] (x1) to node {$-0.08$} (x3);
\draw [corr] (x2) to node {$-0.02$} (x3);
\draw [paths] (e1) to node {$1$} (c);
\draw [paths] (e2) to node {$1$} (s) ;
\end{tikzpicture} 
\caption{A hypothetical diagram created using an R package called \textit{daggity}\cite{R21} with arrows depicting causal relationships between covariates (X1, X2, X3) and outcomes (class and survival) in an observational study setting. Data were simulated using a \textit{simulateSEM} function within \textit{daggity} package that interprets a path diagram, such as that depicted here, using Wright's rules\cite{R29}. Both the class variable (C) and survival (S) are initially simulated as multivariate normal, with C transformed to binary (i.e.\ representing latent subgroups of patients) and S transformed to Weibull as described in the main text.}
\label{fig:DAG}
\end{figure}
A standard one-class Cox proportional hazard model was compared against a two-class Cox proportional hazard model using a heterogeneous population that was simulated according to the hypothetical structure shown in Figure \ref{fig:DAG}. We computed AUC as an index for evaluating the prediction accuracy of the Cox proportional hazard models as proposed by Heagerty et al, $2000$\cite{R24}. \\
We designed our simulation to mimic an observational study setting to enable us to investigate the role of lifecourse exposures in predicting both class-membership and the health outcome (e.g.\ survival). We let 
$\textbf{X}=\{X_1, X_2, X_3\}$ be a vector of covariates drawn from the multivariate normal distribution.  The class $C$ variable was generated by arbitrarily splitting a normally distributed variable into two groups:  $70\%$ in class $1$ and  $30\%$ in class $2$ as approximately observed in the real-world dataset. \\ The Weibull distribution (i.e.\  $\mathcal{W}(\eta, \lambda)$) characterised by two parameters, $\eta$ (shape parameter) and 
$\lambda$ (scale parameter) is frequently assumed for the distribution of the survival times in most studies \cite{R27}. The time to event, $S$ was generated from a normal (i.e\  $\mathcal{N}(\mu,\sigma^2)$) and converted to a Weibull random variable, $T$  using the following transformation \cite{R28}:
\begin{equation}\label{eq:6}
\displaystyle{T = \eta \Bigg\{-\ln\Bigg( \frac{1}{2}\bigg[1-\textit{erf} \Big(\frac{s-\mu_s}{\sigma_s\sqrt{2}}\Big)\bigg]\Bigg)\Bigg\}^{\frac{1}{\lambda}}}
\end{equation}
where 
\begin{equation}\label{eq:7}
\displaystyle{\textit{erf} \Bigg(\frac{s-\mu_s}{\sigma_s \sqrt{2}}\Bigg) = \frac{2}{\sqrt{\pi}} \int_0^s \exp (-t^2) dt}.
\end{equation}
is an error function. \\   \\
The process uses the cumulative distribution function (CDF) to convert data from $\mathcal{N}(\mu,\sigma^2)$ to $\mathcal{W}(\eta, \lambda)$. Assuming that the CDF for a normal and Weibull are defined as in equations \ref{eq:8} and \ref{eq:9}.
\begin{equation}\label{eq:8}
\displaystyle{F(s) =\frac{1}{2}\bigg[1+\textit{erf} \Big(\frac{s-\mu_s}{\sigma_s\sqrt{2}}\Big)\bigg]},
\end{equation}
\begin{equation}\label{eq:9}
\displaystyle{F(t) = 1-\exp \bigg(-\bigg(\frac{t}{\eta}\bigg)^{\lambda}\bigg)},
\end{equation} \\
Normally distributed data points are first converted to a standard uniform through a CDF transformation described in equation \ref{eq:8} and finally converted to a Weibull CDF shown in equation \ref{eq:9} with specified parameters. Equating $F(s)$ to $F(t)$, the transformed time to event variable, $S$ can be expressed as in equation \ref{eq:6}.  \\
In our simulation,  the shape parameter, $\eta$ was set at $0.5$ to yield a distribution with  a decreasing failure rate over time while the scale parameter, $\lambda$ was set at $1$ to yield an exponential distribution which is a special case of the Weibull.  \\
\noindent
We hypothesized a lifecourse scenario within an observational study setting in which $X_1$, $X_2$ and $X_3$ are causal factors measured at different times during an individual's lifecourse such that we arbitrarily assume $X_1$ (e.g.\ genetics, birth weight, early-life nutritional health) is an exposure measured during the early life and $X_2$ (e.g.\ obesity, lack of physical exercise, smoking) occurs midway in the lifecourse before  $X_3$ (e.g.\ drug adherence), which occurs closest to the outcome being predicted.  \\
We simulated $100$ datasets with a sample size and censoring fixed at $1800$ and $30\%$ respectively informed by real-world data that was used for the initial analysis. We specified the path coefficients between the three covariates and outcomes (i.e\ class and survival) to quantify the causal relationships proposed. The covariances between covariates were specified using the double sided arrows as shown in Figure \ref{fig:DAG}. The implied correlation matrix, $\Omega$ for the theoretical path model in Figure \ref{fig:DAG} with variables $C$, $S$, $X1$, $X2$, $X3$ is given below: 
\[\Omega =
\begin{bmatrix}
1.00 & 0.73 & 0.69 & -0.85 & -0.18 \\
0.73 & 1.00 & 0.17 &  -0.97 & 0.54 \\
0.69 & 0.17 & 1.00 & -0.33 & -0.62 \\
-0.85 & -0.97 & -0.33 &  1.00 & -0.37 \\
-0.18 & 0.54 & -0.62 & -0.37 &  1.00
\end{bmatrix}
\]
The variable $X_1$ was assumed to have a strong causal effect on $C$ (i.e.\ $X_1$ was assumed to be a strong cause of class-membership) and its causal effect on $S$ was entirely mediated via $C$- this depicts how early life factors, such as genetics and developmental factors, may strongly determine later-life subgroup differences in cardiovascular disease risk. $X2$ was assumed to have a causal impact on both $C$ and $S$- this depicts how lifestyle factors, such as differences in physical exercise, may modify any underlying genetic predisposition to cardiovascular disease risk. The covariate $X3$ was assumed to have a strong causal effect on $S$ and no causal effect on $C$- this depicts how response to the diagnosis of disease, such as adherence to treatment regimes, influence subsequent survival risk. \\
Once the two multivariate normal variables for $C$ and $S$ were transformed into binary and Weibull distributed, respectively, we fitted the standard Cox proportional hazards model to evaluate the predictive performance of the model, calculating AUC. We also explored the two-class model by fitting the Cox proportional hazards model within a latent class framework and calculating AUC. Both evaluations were repeated for all $100$ simulated datasets. 
\subsection{Simulation results}
\begin{figure}[th]
  \centering
    \includegraphics[scale=0.40]{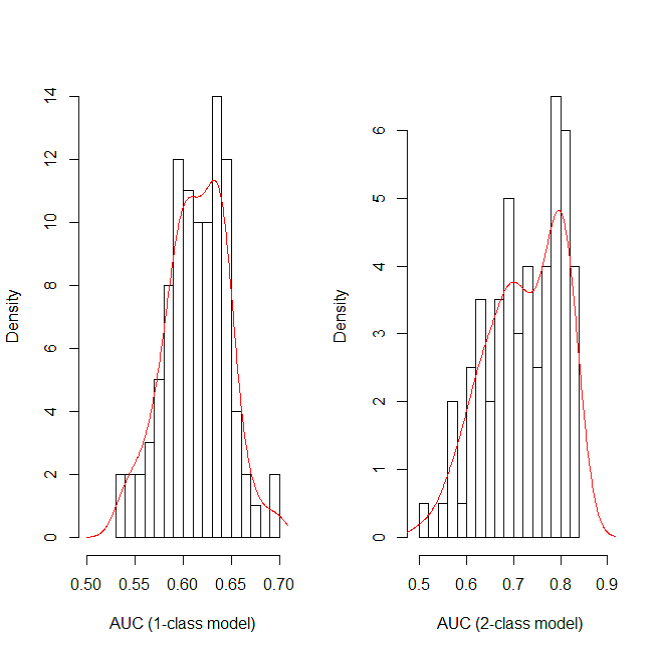}
    \caption{Kernel density plots showing the distribution of AUC for the 100 simulations. } 
    \label{fig:kernel}
 \end{figure}

The underlying correlation matrix, $\Omega ^{\prime}$ for the simulated dataset with variables $C$, $S$, $X1$, $X2$, $X3$ is given below: 
\[\Omega ^{\prime} =
\begin{bmatrix}
1.00 & 0.42 & 0.36 & -0.24 & 0.00 \\
0.42 & 1.00 & 0.16 &  -0.38 & 0.38 \\
0.36 & 0.16 & 1.00 & 0.02 & -0.08 \\
-0.24 & -0.38 & 0.02 &  1.00 & -0.05 \\
0.00 & 0.38 & -0.08 & -0.05 &  1.00
\end{bmatrix}
\]
The correlation matrix for the simulated data fulfils the goal of our simulation which is to yield a dataset in which our assumed causal structure is realistic.  
Results from the simulation study in Table \ref{tab:simulations}, shows that our proposed latent class modelling framework improves the prediction of survival times. The mean AUC increases from an average of $0.61$ in the one-class model to $0.72$ in the two-class model. The distribution of AUC also depicts an overall improvement between the models. The mean and median of AUC are almost equal as shown in the  kernel density plots for both one-class and two-class models in Figure \ref{fig:kernel}. 
\begin{table} [th]
\small\sf
\caption {Summary area under the ROC curves to assess the predictive performance of two modelling frameworks for the 100 simulations  } \label{tab:simulations}   \centering
\begin{tabular}{ l  l  l l}
\toprule
%\hline
$\textbf{Model}$ & $\textbf{Mean AUC$ $(SD)}$ & $\textbf{Median AUC$ $(IQR)}$  \\  
\midrule
$\text{One-Class}$ & $0.63$ $(0.03)$ & $0.63$ $(0.61-0.65)$   \\  
$\text{Two-class}$ & $0.75$ $(0.06)$ & $0.75$ $(0.71-0.80)$  \\   
\bottomrule
\end{tabular} 
\begin{tablenotes}
\centering
      \scriptsize
 \item \textbf{${}^\star$  AUC= Area under the Curve}
 \item  \textbf{${}^\star$ IQR= Inter Quartile Range}
  \item \textbf{${}^\star$ SD =Standard deviation}
    \end{tablenotes}
\end{table} 
 \newpage
\section{Discussion}
In this paper, we have proposed the use of a semi-parametric Cox proportional hazard (CPH) model within a latent class framework for prediction of survival time. This model provides an improvement to the standard CPH model in which a single model is used to predict survival time for the whole population. In the latent class modelling framework, the model uses covariates to identify patients with similar risk and forms mutually exclusive subgroups. Each subgroup is then estimated by its own risk model, thereby improving prediction over standard CPH model. By including covariates to predict class memberships, this method allows for the separation of patients into homogeneous subgroups. By estimating different regression models within each class, this method optimally minimises the mean square error and provides more accurate estimates that are tailored for each subgroup, thereby improving both subgroup and population prediction. \\ 
The present analysis was primarily aimed at comparing the predictive accuracy of our proposed modelling framework against the standard method. We conducted a simulation study to assess the efficacy of our proposed model using 100 simulated datasets. In both real and simulated cases, our model has successfully outperformed the standard approach in terms of prediction. There are clear advantages that our proposed methodology demonstrates over the standard approach. We note that by grouping patients with similar characteristics into latent classes, and estimating a different model for each class, our method acknowledges any heterogeneity that is within a population. Predictions based on a single model for the whole population fail to account for this inherent heterogeneity that is likely present within the population. Modelling within a latent class framework would also help researchers to evaluate which subgroup of patients might need different treatment options through the identification of different risk profiles in each class.\\
Classifying predictors according to their possible causal impact within a lifecourse framework would also help researchers to understand which risk factors are likely to better predict either class membership, the outcome in each subgroup of patients, or both.  \\ 
Although we set out to improve prediction, not to evaluate causality, by thinking of predictors within a causal framework it is possible to understand the potential role of predictors in the model by classifying them according to their relevance to the lifecourse. For instance, early-life predictors encapsulate factors such as genetics, birth weight (that reflects in-utero programming), or childhood, adolescent and early adulthood lifestyle factors that set in-train differences between individuals such that population-level subgroups emerge in early adulthood. Mid-life predictors encapsulate factors such as lack of physical exercise, being overweight, and smoking etc.\ that cements inherent subgroup differences in the population in terms of health risk profile and may impact the response to medication or other (e.g. lifestyle) interventions. Contemporaneous predictors encapsulate factors acting on a health condition right up to the event of death, such as adherence to medication or the treatment modality directly. The distinction amongst predictors framed in this way helps researchers to understand which are likely to be better at predicting class membership or the outcome event, or both. \\
It is important that researchers avoid making any causal interpretations from the prediction model. The model coefficients cannot be interpreted directly and estimates of causal impact cannot be inferred, despite potentially employing causal knowledge in understanding the role of model predictors. In-order to make any meaningful causal interpretations, researchers are required to employ appropriate causal inference methodology, which involves identifying a suitable set of covariates that minimize bias due to confounding. One way of achieving this is to use a directed acyclic graph (DAG) which can be done in R using the package \textit{daggity}\cite{R21}. DAGs offer a systematic way of identifying an appropriate set of model covariates from which confounder adjustment can be made to achieve robust causal inference if all known or anticipated confounding is recorded in the dataset. \\
Based on the results presented in this paper, there is strong evidence that our proposed new method can provide better predictions and the role of predictors can have clearer meaning in terms of their occurrence within a lifecourse framework. However, it is worth noting that our proposed modelling framework is affected by the presence of missing data which reduces the prediction accuracy of the model.
Future research must explore the impact of missing data and the role for missing data methods to mitigate reduced prediction accuracy, along with development of guidelines on how to implement these models in a range of software packages. Where it is found that machine learning methods\cite{R22, R23}(e.g.\ artificial neural networks) outperform standard statistical regression models, further work is also warranted to integrate the proposed latent class approach with machine learning techniques, as appropriate. Further work should then extend to the development of new packages in R, specifically for this purpose. \\
Overall, our proposed method presents a positive improvement to the standard one-size-fits-all approach that is common in the literature for prediction modelling, and we recommend researchers to adopt this new approach as it offers a powerful way of assessing risk in patient outcomes by identifying subgroups that likely exist within most large and complex health data. 
\subsection{Acknowledgements} 
The author(s) would like to thank the Commonwealth Scholarship Commission and all participants in the UK-HEART 2 study. 

\subsection{Declaration}
The author(s) declared no potential conflicts of interest with respect to authorship, and/or publication of this article. 

\subsection{Funding}
John L. Mbotwa is funded by the Commonwealth Scholarship Commission and this work is part of his PhD at the University of Leeds. However, the author(s) received no financial support for the authorship, and/or publication of this article.  

\end{document}